\documentclass[aps,prl,twocolumn,notitlepage,superscriptaddress,longbibliography]{revtex4-1}

\usepackage{graphicx}
\usepackage{dcolumn}
\usepackage{bm}
\usepackage{amssymb}
\usepackage{microtype}
\usepackage{xfrac}
\usepackage{makecell}
\usepackage{array}
\usepackage[version=4]{mhchem}
\usepackage{gensymb}
\usepackage{multirow}
\usepackage{physics}
\usepackage[colorlinks=true]{hyperref}
\usepackage{physics}
\usepackage{bm}

\begin{document}

\title{Tuning the Hall response of a non-collinear antiferromagnet with spin-transfer torques and oscillating magnetic fields}
\author{Sayak Dasgupta}
\email{sayak.dasgupta@ubc.ca}
\affiliation{Department of Physics and Astronomy $\&$ Stewart Blusson Quantum Matter Institute, University of British Columbia, Vancouver,  British Columbia V6T 1Z1, Canada}
\affiliation{Institute for Solid State Physics, University of Tokyo, Kashiwa 277-8581, Japan}
\author{Oleg A. Tretiakov}
\email{o.tretiakov@unsw.edu.au}
\affiliation{School of Physics, The University of New South Wales, Sydney 2052, Australia}

\begin{abstract}
The kagome lattice antiferromagnets Mn$_3$X(= Sn, Ge) have a non-collinear 120$^\circ$ ordered ground state, which engenders a strong anomalous Hall response. It has been shown that this response is linked to the magnetic order and can be manipulated through it. Here we use a combination of strain and spin-transfer torques to control the magnetic order and hence switch deterministically between states of different chirality. Each of these chiral ground states has an anomalous Hall conductivity tensor in a different direction. Furthermore, we show that a similar manipulation of the strained sample can be obtained through oscillating magnetic fields, potentially opening a pathway to optical switching in these materials.
\end{abstract}
\maketitle

A significant direction of current spintronics research lies in characterizing and understanding the anomalous Hall (AH) response of antiferromagnets. Contrary to conventional wisdom, which suggests a proportionality between the Hall signal and the magnetization of the system, these systems show large AH responses and tiny induced magnetic moments. It is now understood that the AH response stems from the electronic structure, especially Weyl points near the Fermi energy \cite{Liu:2017}. In particular focus are the kagome-lattice based magnets Mn$_3$X, which show very high AH signals at room temperature with almost negligible induced magnetic moments. The Hall response is an intrinsic property of the anti-chiral 120$^\circ$ order, which exists in the range $T = 5-380$ K in Mn$_3$Ge and $T = 50-420$ K in Mn$_3$Sn \cite{Nakatsuji2015,nakatsujimn_3ge,nayak-hall-mn3ge}.

\begin{figure}[h!]
		\centering
		\includegraphics[width = \columnwidth]{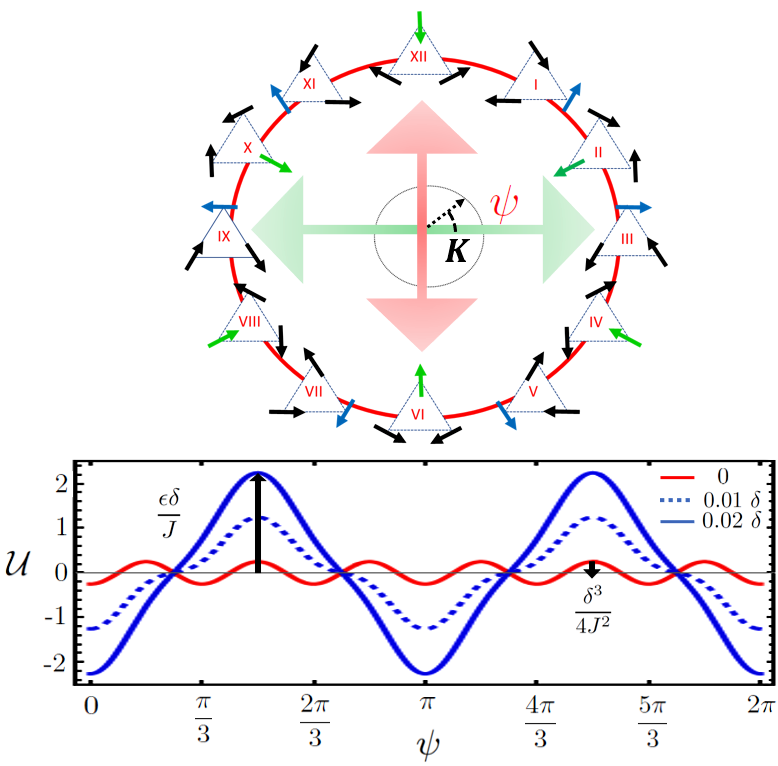}
		\caption{Magnetically ordered states of Mn$_3$Sn (blue easy spin) and Mn$_3$Ge (green easy spin) are shown with $\bf K$. The direction of $\mathbf{K}$ is given by local mirror planes, it is parallel to the spin that points along the local easy axis for Mn$_3$Sn and opposite for Mn$_3$Ge. The six ground states for Mn$_3$Sn are shown in red in the free energy plot. Strain distorts the energy landscape such that as $\epsilon \to \delta$ the local symmetry changes to $C_2$ in the kagome plane. A positive (compressive) strain along the $y$-axis (red double arrow) stabilizes states at $\psi = 0, \pi$.}
		\label{fig.geometry-modes}
\end{figure}

The 120$^\circ$ state can be expressed through the normal modes belonging to the irreducible representation of the $D_{3h}$ symmetry group. These are grouped into modes in the kagome plane -- $\alpha_0$ and the doublets $\bm{\alpha} = (\alpha_x,\alpha_y)$, and out-of-plane $\beta_0$ and $\bm{\beta} = (\beta_x,\beta_y)$. The two doublets $\bm{\alpha}$ and $\bm{\beta}$ transform as vectors in the kagome plane \cite{Mineev1996,dasgupta-mn3ge-2020}. The ground state manifold comprises states that lie on the hands of a clock, with even hours for Mn$_3$Ge and odd hours for Mn$_3$Sn, enforced by an easy-axis anisotropy~\cite{Chen-mn3ge-2020,dasgupta-mn3ge-2020,sayak-future}. 

The singlet $\alpha_0$, which represents uniform rotation of all the spins in the kagome plane forms a new order parameter $\mathbf{K} = (\cos\psi,\sin\psi)$ with $\psi = \alpha_0/\sqrt{3}$. Its orientation is given by the single spin that satisfies the easy axis in each of the clock states, see Fig.~\ref{fig.geometry-modes}. This order parameter, $\mathbf{K}$, couples to the electronic structure via the local spins, leading to a Hall conductivity tensor proportional to $\mathbf{K}$, i.e. $\sigma^{H}_{ij} = (e^2/2 \pi h) \zeta \epsilon^{ijk}\mathbf{K}$ where $\zeta$ is given by the electronic band structure \cite{Liu:2017}. 

Thus by controlling the local spin order we can manipulate the \textit{orientation} of the Hall vector in the kagome plane. This was achieved using an uniaxial strain in a constant magnetic field \cite{ikhlas-future}. There a uniaxial strain in the kagome plane changes the local $C_6$ symmetry to a $C_2$, see Fig.~\ref{fig.geometry-modes}. The magnetic order parameter, and hence the Hall vector, responds to this new symmetry aligning along an axis chosen by the strain \cite{ikhlas-future,sayak-future}. 

In this paper, we present two ways of manipulating $\mathbf{K}$ in a strained sample: (1) with an oscillating magnetic field and (2) with spin-transfer torques (STT). The former is of interest in optical experiments such as \cite{Disa2020}, where THz pulses have been used to switch the order in a two-sublattice antiferromagnet. The latter is partly motivated by the manipulation of the local spin order through an STT achieved in \cite{Takeuchi2021,Tsai2020}. We show that by augmenting the setup with a strain, we can use it to control the direction of the Hall vector with great precision. 

The implication of strain to control the spin-wave spectrum in two-sublattice antiferromagnets has been studied in \cite{kittel58,zhang-su3,dasgupta21}, here we investigate a three-sublattice system. Notably, we use strain to control the order parameter $\mathbf{K}$. To do so we need to exert strains large enough to overcome the small uniaxial anisotropy in these systems. The required strains are $\sim 0.1\%$ of the exchange energy $J$. Such strains are considerably smaller than what is required to affect the electronic band structure ($\sim 1 \%$ of $J$). In each of the allowed clock states, see Fig.~\ref{fig.geometry-modes}, the system has an AH response of the same size but in different directions. Recently, there has been extensive experimental work done on Mn$_3$X systems showing the manipulation of the AH effect through strain variations. For instance in \cite{wang2019integration,guo2020giant}, where epitaxial strains are used to effect large changes in the anomalous Hall responses of  Mn$_3$Sn and Mn$_3$Ga respectively. Strain has also been used very recently to reverse the sign of the Hall response completely in a constant magnetic field \cite{ikhlas-future}.

Additionally, we investigate the dynamics of the soft modes $\psi$ and $\bm{\beta}$ under an oscillating magnetic field and spin current. In the process, we show that strain and time-dependent magnetic fields can be used to elicit a wide range of antiferromagnetic resonances, which might be of importance for future experiments and devices designs. In all our numerical simulations we set the value of the effective exchange constant $J = 1$ meV and measure all other energies with respect to $J$. This sets a natural frequency scale of $3.3$ THz and time scale $t = 0.3$ ps, which is the unit of time in all our results. 

\textbf{Energy Functional:} The magnetic energy functional for Mn$_3$X can be captured by the minimal model \cite{Chen-mn3ge-2020,boothroyd-mn3ge-2020,chaudhary2022}:
\begin{equation}
    \label{eq.minimal-H}
    \mathcal{H} = \sum_{ij} J_{ij} \mathbf{S}_i\cdot\mathbf{S}_j + D\sum_{<i,j>} \mathbf{\hat{z}}\cdot(\mathbf{S}_i \times \mathbf{S}_j)
    - \delta \sum_i (\mathbf{n}_i\cdot\mathbf{S}_i)^2,
\end{equation}
where the first term describes the exchange interaction, the second describes the Dzyaloshinskii-Moriya (DM) \cite{DZYALOSHINSKY1958241,Moriya1960} interaction with the out-of-plane DM vector, and the third gives the magnetic anisotropy. The exchange interaction expanded near the $\Gamma$ point to quadratic order in soft and hard modes takes the form:
\begin{eqnarray}
    \label{eq.excahnge-single-layer}
    \mathcal{U}_{J} = J \sum_{<ij>}\mathbf{S}_i\cdot\mathbf{S}_j = \frac{3 J}{2} S^2 (\bm{\alpha}\cdot\bm{\alpha} + 2\beta_0^2).
\end{eqnarray}
Modes $\beta_0$ and $\bm{\alpha}$, which induce a net magnetization, are penalized by exchange interaction and can be integrated out to generate inertia for the soft modes $\psi$ and $\bm{\beta}$, as in \cite{dasgupta-mn3ge-2020}. This leads to the kinetic term:
\begin{equation}
    \mathcal{K} = \frac{\rho_{\psi}}{2} \dot{\psi}^2 + \frac{\rho_{\beta_x}}{2} \dot{\beta}_x^2 + \frac{\rho_{\beta_y}}{2} \dot{\beta}_y^2,
\end{equation}
where $\rho_\psi = 1/(2 J) = (3/2)\rho_{\beta}$, with $J = J_1 + J_2$. The remaining interactions from the DM vector, and the local anisotropy form an energy functional in terms of the soft modes. This is modified by an in-plane uniaxial strain $(\epsilon_{xx}-\epsilon_{yy},2\epsilon_{xy}) = \epsilon (\cos 2\psi_\epsilon,\sin 2\psi_\epsilon)$, where $2\epsilon_{ij} = \partial_i u_j + \partial_j u_i$, with $\mathbf{u}$ being lattice displacements.

The effect of strain in the Hamiltonian in Eq.~(\ref{eq.minimal-H}) is captured through the variation of the Heisenberg exchange with lattice site displacements $\sum_{ij} \left[ (\partial J / \partial \mathbf{u}) \cdot \delta\mathbf{u}\right]\mathbf{S}_i\cdot\mathbf{S}_j$, following \cite{tchernyshyov2002-prl,tchernyshyov2002,dasgupta21}. The exact form of the decay of $J_{ij}$ with separation is not important and we retain only the first derivative correction. This correction is substantial in Mn$_3$X as evident from the very strong magnon phonon coupling in the antichiral 120$^\circ$ phase seen and calculated in \cite{Chen-mn3ge-2020} and also estimated in \cite{sukhanov2018} through measurement of magnetic order under pressure. The total energy functional to quadratic order in soft modes:
\begin{widetext}
\begin{eqnarray}
    \label{eq.full-functional}  
    \mathcal{U} &=& \left(\sqrt{3} D + \frac{\delta}{2}\right)\bm{\beta}\cdot\bm{\beta} - \frac{\delta}{4} \left[ (\beta_x^2 - \beta_y^2) \cos 2\psi - 2\beta_x\beta_y \sin 2\psi \right]  - \frac{\delta^3}{4 J^2} \cos 6\psi \nonumber \\
    &&+ \frac{\epsilon}{4} \left[ (\beta_x^2 - \beta_y^2)\cos 2\psi_\epsilon + 2\beta_x\beta_y\sin 2\psi_\epsilon \right] + \frac{\delta\epsilon}{J}\cos 2(\psi + \psi_\epsilon) - \frac{\delta\epsilon}{4J^2}[\epsilon\cos (2\psi - 4\psi_\epsilon) - 2\delta\cos(4\psi - 2\psi_\epsilon)], 
\end{eqnarray}
\end{widetext}
where we have absorbed the factor $3/2$ into $\epsilon$. The small pinning energy of the $\psi$ mode $\propto\sqrt{\delta^3/J}$ implies that one can easily affect the dynamics of the $\psi$ mode.

\begin{figure}[htb]
		\centering
		\includegraphics[width = \columnwidth]{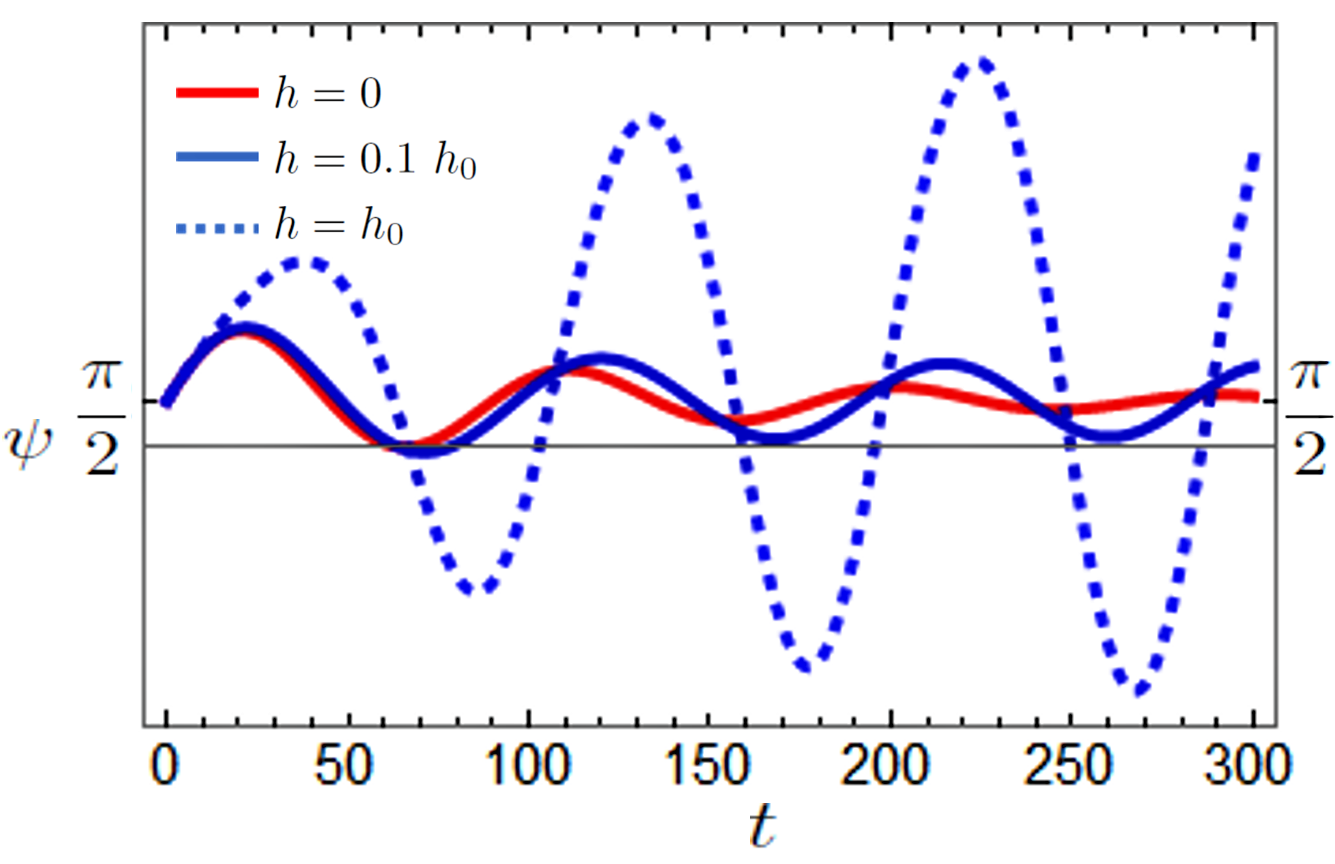}
		\caption{Oscillation around ground state at $\psi = \pi/2$ and initial angular velocity $\dot{\psi}(0) = 0.01$, with a fixed strain angle $\psi_\epsilon = \pi/2$. We set the exchange $J=1$ and anisotropy strength $\delta = 0.01\, J$, strain is set to $\epsilon = 5\, \delta$. The oscillating component $h$ is varied and we can see the amplitude growth. The damping constant is $\alpha_D = 0.01$. }
		\label{fig.osci-mag-ground}
\end{figure}

\textbf{Time varying magnetic field:}
We now look at the precessional dynamics of the $\psi$ mode under a magnetic field:
\begin{equation}
\label{eq.mag_field_drive}
    \mathbf{H} = h_0(1,0,0) + h(\cos\nu t, \sin\nu t,0),
\end{equation}
where the direction of the constant field is chosen to simplify the expressions. We absorb the gyromagnetic ratio $\gamma$ into the field strength and set spin length $S = 1$. To this we apply a strain along the $y$-axis, $\psi_\epsilon = \pi/2$. Let us first analyze the case $h = 0$. The energy terms we retain are of the order $1/J$:
\begin{equation}
\label{Umag}
    \mathcal{U}_{mag} = - \frac{1}{J} \left[ h_0(\delta - \epsilon)\sin\psi + \delta\epsilon \cos(2\psi) \right].
\end{equation}
All other terms are highly suppressed by the exchange energy scale and do not contribute to the dynamics. From Eq.~(\ref{Umag}) it is clear that the dynamics of the $\psi$ mode is now controlled by the $C_2$ anisotropy coming from the strain. This blurs the distinction between the Sn and Ge compounds and we can traverse the clock manifold continuously using the appropriate size and sign of strain. Note that the six-fold anisotropy $\propto \delta^3/J^2$ is still present, but its contribution to stabilizing a ground state is negligible if $\epsilon\simeq\delta$. The equation of motion around the ground state $\psi = \pi/2$ is
\begin{equation}
    \rho_{\psi}\Ddot{\psi} = \frac{1}{J}[h_0(\epsilon - \delta) + 4 \delta\epsilon]\left(\psi - \frac{\pi}{2}\right) - \alpha_D\dot{\psi},
\end{equation}
where $\alpha_D$ is the damping. Small perturbations around the state result in decaying oscillations, Fig.~\ref{fig.osci-mag-ground}. The natural frequency around this ground state $\nu_{\pi/2} \propto \sqrt{h_0(\epsilon - \delta) + 4\delta\epsilon}$ (see Table~\ref{table.nat_f} for the others). This can be tuned by changing the orientation of strain, $\psi_{\epsilon}$, and $\mathbf{H}$.

\begin{table}[h!]
\begin{center}
\begin{tabular}{|c|c|} 
\hline
   $\psi$  & $J\rho\nu_0^2$  \\ \hline
     0     &   $-4\delta\epsilon$ \\ \hline
    $\frac{\pi}{6}$ &  $\frac{1}{2}[h_0(\epsilon - \delta) - 4 \delta\epsilon]$ \\ \hline
    $\frac{\pi}{3}$ &  $\frac{1}{2}[\sqrt{3}h_0(\epsilon - \delta) + 4 \delta\epsilon]$ \\ \hline
    $\frac{\pi}{2}$ &  $h_0(\epsilon - \delta) + 4\delta\epsilon$ \\
\hline
\end{tabular}
\caption{Natural frequencies at $\psi_\epsilon = \pi/2$ and a constant magnetic field $h_0$ along $x$-axis.}
\label{table.nat_f}
\end{center}
\end{table}

Let us now turn on the oscillating field. Now, since the dynamics is that of a forced oscillator we can tune the frequency close to or away from the natural values, see Table.~\ref{table.nat_f}. For a low enough dissipation this is close to the resonant frequency. The profile attains the expected growth at $\nu_{\pi/2}$ on increasing the drive strength to $h = 0.1\, h_0$ and to $h = h_0$ in Fig.~\ref{fig.osci-mag-ground}.

\begin{figure}[t]
		\centering
		\includegraphics[width = 0.9 \columnwidth]{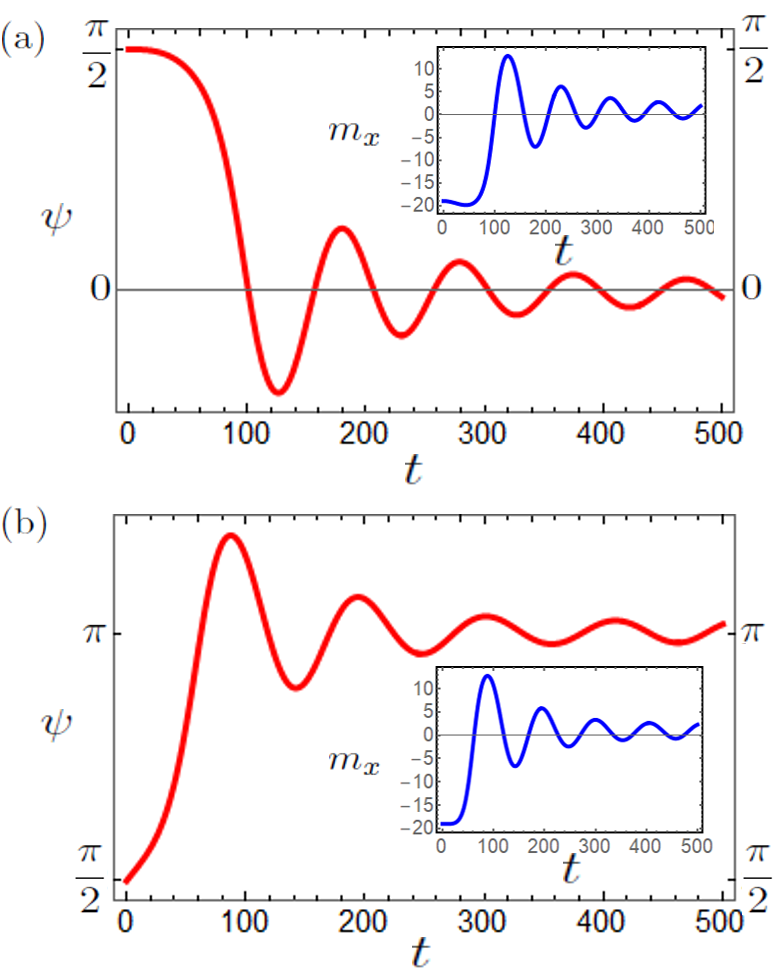}
		\caption{The strain switches sign after the drive is turned on. The magnetic field is $h = 0.1~h_0$ with $|h_0| = 0.1 ~\delta$ in (a) and $|h_0| = -0.1~ \delta$ in (b). Initial angular velocity is set to $\dot{\psi}(0) = 0$. The two states saturate to a $\psi$ perpendicular to the initial state, i.e., from $K_y \to \pm K_x$. This switches the Hall conductivity from $\sigma^{xz}$ to $\pm \sigma^{yz}$ or a magnitude switch from $|\sigma|_{xz} = 1$ to $|\sigma|_{xz} = 0$. In the insets the magnetization $m_x$ is measured in $10^{-3}\mu_B$ units.}
		\label{fig.osci-mag-switch}
\end{figure}

In this protocol, away from resonant growth, the Hall angle $\psi$ saturates to its initial state. We can change this if we switch the \textit{sign} of the strain while the drive is on. A change in the sign of strain lowers the energies on the clock perpendicular to the initial state. Any perturbation delivered by the oscillating component of the field in this switched strain configuration leads to the system settling in the newly favored ground state(s), see Fig.~\ref{fig.osci-mag-switch}. 

\textbf{Adiabatic Spin Transfer Torque (STT):}
\textcolor{black}{For the spin transfer torque injection we use the setup in Takeuchi $et~al$\cite{Takeuchi2021} (see Fig. 2 there), with a spin current being pumped into the kagome plane from below. This is similar to a Spin Orbit Torque setup in that the spin is being injected locally into the sample \cite{go2022noncollinear}}. The adiabatic STT can be incorporated through a Rayleigh term $\mathcal{R}_{STT} = \eta (\mathbf{m}\times\mathbf{m}_0)\cdot\dot{\mathbf{m}}$ where $\mathbf{m}_0$ is the polarization of the spin current. Consider a spin polarization out of the kagome plane $\mathbf{\hat{m}}_0 = \mathbf{\hat{z}}$. The $\psi$ mode responds strongly while the $\bm{\beta}$ doublet responds only to a current polarized in the kagome plane. We turn the magnetic field off and assume that a single domain state is created by cooling down in a magnetic field. The relevant knobs remaining are strain, $\epsilon$, and the STT amplitude, $\eta$. The dynamical equation for the $\psi$ mode with $\psi_\epsilon = \pi/2$ is
\begin{eqnarray}
 \rho_{\psi} \Ddot{\psi} &=& -\frac{2\delta\epsilon}{J}\sin(2\psi) \nonumber \\
&&- \frac{\delta}{2 J^2}[\epsilon^2\sin(2\psi) + 4\delta\epsilon\sin(4\psi) + 3\delta^2\sin(6\psi)] \nonumber \\
&&+ 3\eta(t) - 3\alpha_D\dot{\psi}.
\end{eqnarray}

\begin{figure}[t]
		\centering
		\includegraphics[width = .9\columnwidth]{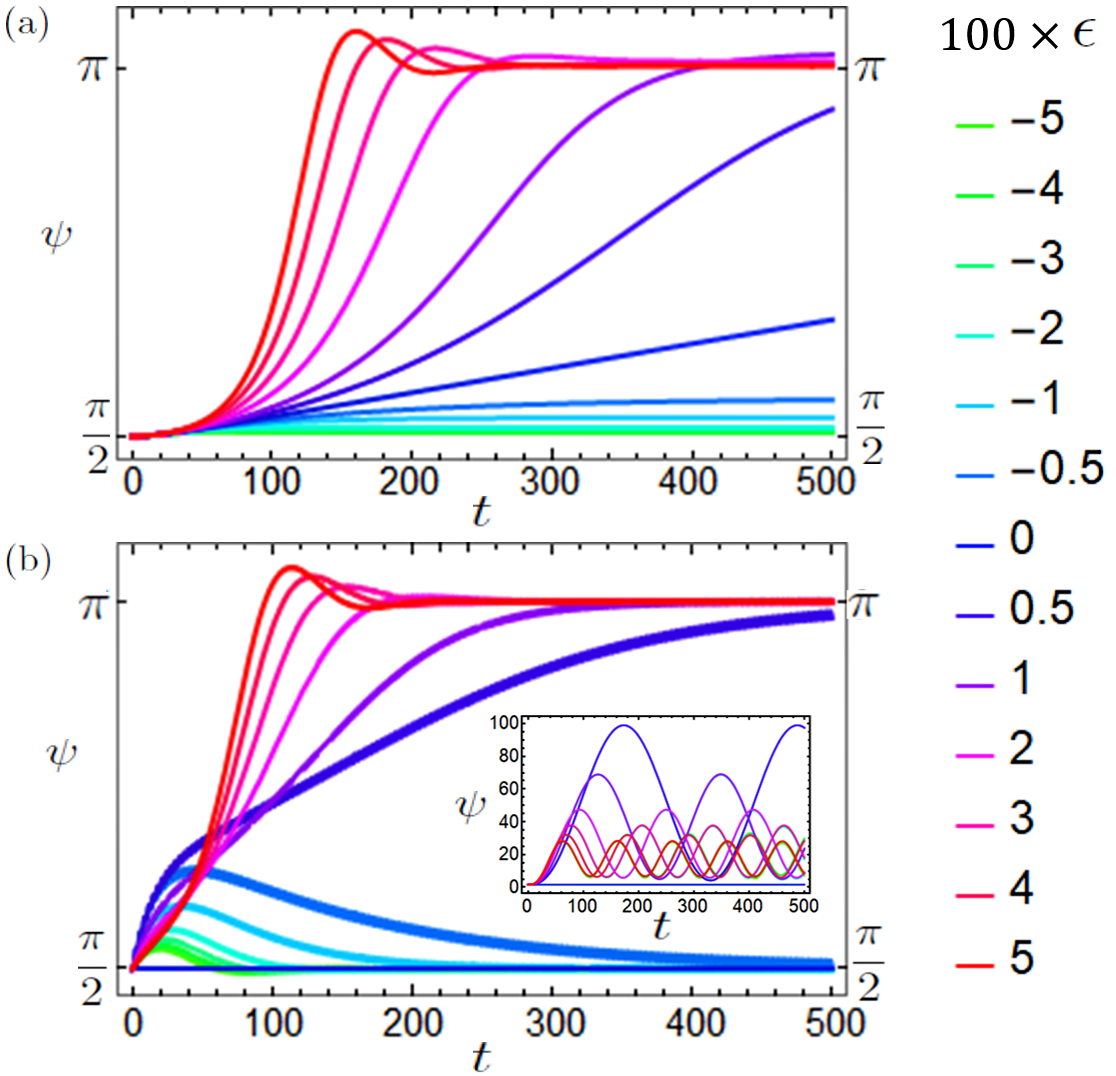}
		\caption{Varying strain at a constant angle $\psi_\epsilon = \pi/2$ and follow the evolution from $\psi(0) = \pi/2$ under an out-of-plane STT. In (a) the STT is constant and small $\eta(t) = \eta_0 = 0.001\,\delta$, in (b) the STT is driven at a frequency much higher than the natural frequency $\eta(t) = \eta_0 \sin(\nu_{\eta} t)$, $\nu_\eta = 100\,\nu_{nat}$, and in the inset at the natural frequency. \textcolor{black}{The $\pi/2$ angle is marked by a black line in (b) to show convergence}. In both cases $\eta_0\simeq\delta$. }
		\label{fig.osci-strain-STT}
\end{figure}

We can play the same switching game as we did with the oscillating magnetic field in Fig.~\ref{fig.osci-mag-switch} here, i.e., starting the evolution of $\psi$ at $\pi/2$ and the strain along the $x$-axis $\psi_\epsilon = 0$ we can adjust the driving amplitude to shift the final orientation of the order parameter, $\psi$. For a very small driving parameter $\eta = 0.001\,\delta$ we can switch from an initial state along $\psi = \pi/2$ to a state $\psi = \pi$ as we tune strain from positive to negative, see Fig.~\ref{fig.osci-strain-STT}(a). Note that in the absence of strain, $\epsilon = 0$, (blue line) we obtain a uniform precession of the $\psi$ mode as observed in \cite{Takeuchi2021}. The perpendicular switch of the state happens sharply as $\epsilon$ switches sign. 

Alternatively, we can use a time varying drive $\eta = \eta \sin(\nu_\eta t)$, and relax the smallness condition on $\eta$. For a fast frequency drive $\nu_\eta \simeq 100\, \nu_{\pi/2}$ the order parameter settles at the minima corresponding to the one favored by the strain anisotropy, at long times even for $\eta \simeq \delta$, see Fig.~\ref{fig.osci-strain-STT}(b). At this value of $\eta$ a driving frequency matching the natural frequency scale of the minima produces a time-dependent precession of the $\psi$ mode, which washes out the Hall effect and magnetization signals as shown in the inset of Fig.~\ref{fig.osci-strain-STT}(b).

\begin{figure}[htb]
		\centering
		\includegraphics[width = 0.96\columnwidth]{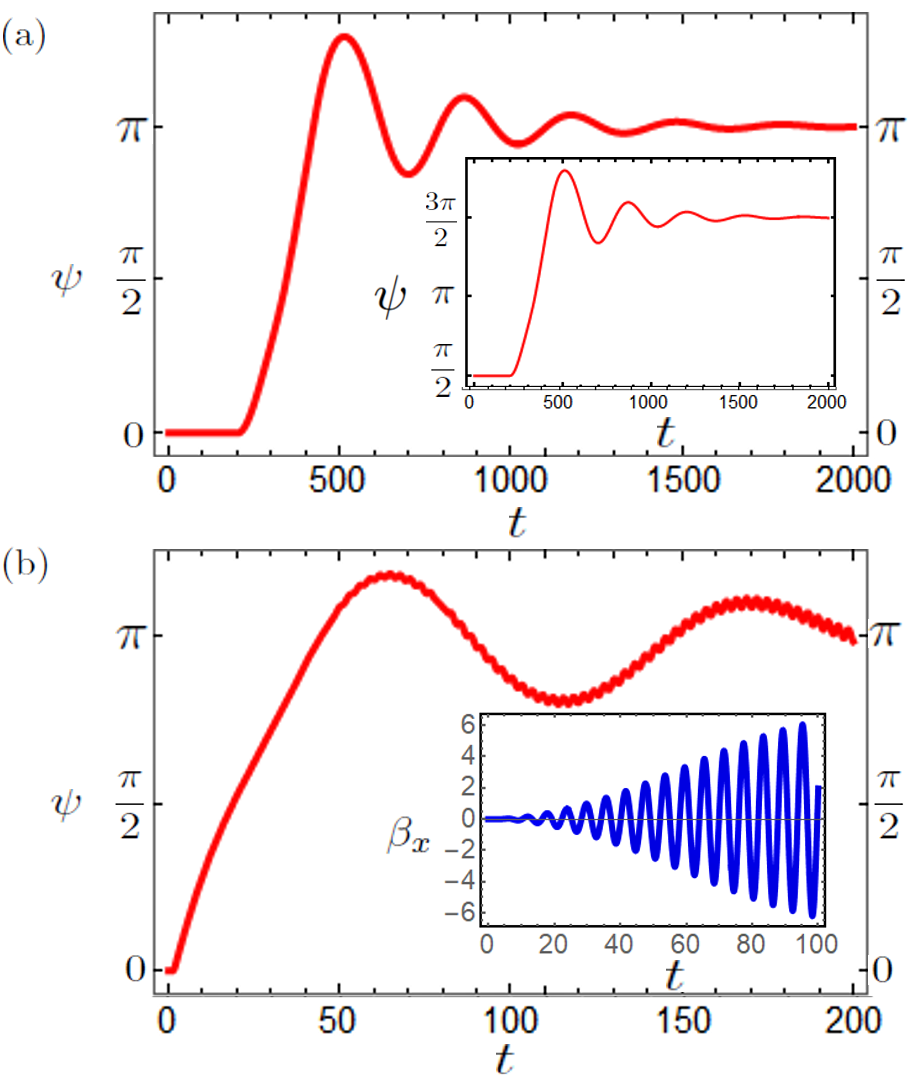}
		\caption{In (a) we use an out-of-plane polarized STT of constant magnitude $\eta = 0.015\, \delta$ and switch from an initial state of $\psi(0) = 0$ ($= \pi/2$ in the inset) to its diametric opposite. A positive strain is applied along the $x$-axis, $\psi_{\epsilon} = 0$, ($= \pi/2$ in the inset) with magnitude $\epsilon = \delta/2$. The drive starts at $t = 200$ and lasts until $t=350$ ($\simeq 45$ ps width), with $\dot{\psi}(0) = 0$. In (b) the coupled system from Eq.~(\ref{eq.STT-couple}) is driven at the natural frequency of $\beta_x$. We initiate the drive using an out-of-plane STT $\eta'(t) = \eta [\Theta(t-1) - \Theta(t-2)]$, where $\Theta(t)$ is the Heaviside step function. The parameters are $\epsilon = - 5\, \delta$, $\eta = 2\, \delta$, and $D = 10\, \delta$ with $\delta = J/100$. The damping constant is set to $\alpha_D = 0.003$.}
		\label{fig.switch-STT}
\end{figure}

\textbf{Full switching with STT :}
In the presence of the STT we can also fully switch the Hall vector at a constant strain through a protocol design. We first settle on a ground state using a strain, say state III in Fig.~\ref{fig.geometry-modes}. Note that on the clock (Fig.~\ref{fig.geometry-modes}) the strain stabilizes diametrically opposite states with reversed signs for the Hall signal, so state IX is also a global minimum of the free energy. We then turn on an out-of-plane polarized STT. This causes $\psi$ mode to precess for a large enough STT magnitude. We let $\psi$ evolve until we cross the state at XII (or VI since both paths are equally probable) and then turn off the spin current. The strain then forces the final state to be diametrically opposite to the initial state, reversing the sign of the Hall signal. This is shown for selected parameters in Fig.~\ref{fig.switch-STT}(a) and can be adjusted according to experimental situations. We can tune both the length of the drive and the strength to achieve the switching.

For a current polarized along the kagome plane, the three modes, $\psi$ and $\bm{\beta}$, are mixed. We analyze this case now with $\mathbf{m}_0 = \sqrt{2/3}\eta\,(1,0,0)$. To keep the analysis tractable we assume that the strain is large enough, $\epsilon > \delta$, to maintain an effective $C_2$ symmetry. The Rayleigh dissipation function in this case reads:
\begin{equation}
\label{eq.dissipation-STT-pol-in}
    \mathcal{R} = \eta \dot{\beta_y} + \eta(\dot{\beta}_x\psi - \dot{\psi}\beta_x) + \frac{\alpha_D}{2}(\dot{\bm{\beta}}\cdot\dot{\bm{\beta}} + 3\dot{\psi}^2).
\end{equation}
We assume that the three modes have a well-defined inertia set by the exchange, with minor modifications from the strain and local anisotropy, which can be ignored. The equations of motion to linear order in $1/J$ are:
\begin{subequations}
\begin{eqnarray} 
\label{eq.STT-couples_psi_beta} 
\!\!\!\!\!\!&&\rho' \Ddot{\psi} = \frac{2\delta\epsilon}{J}\sin[2(\psi + \psi_{\epsilon})] + \eta \beta_x - 3\alpha_D\dot{\psi} + 3\eta'(t), \\
\label{eq.STT-couples_psi_beta2} 
\!\!\!\!\!\!&&\frac{2\rho'}{3}\Ddot{\beta}_x = \left(\frac{\epsilon}{2} - 2\sqrt{3}D\right)\beta_x - \eta \psi - \alpha_D\dot{\beta}_x,  \\ 
\!\!\!\!\!\!&&\frac{2\rho'}{3}\Ddot{\beta}_y = -\left(\frac{\epsilon}{2} + 2\sqrt{3}D\right)\beta_y  - \eta - \alpha_D\dot{\beta}_y, 
\end{eqnarray}
\label{eq.STT-couple}
\end{subequations}
where we have added an STT with an out-of-plane polarization to initiate dynamics $\eta' \simeq \eta$ through a step function.

From this we can see that $\beta_y$ is decoupled in this configuration and driven by the STT term. The $\psi$ and $\beta_x$ modes are coupled through the spin torque amplitude and the coupled system is not forced, and the coupling is purely through the STT. Note that, if we had chosen a spin current polarization along $\mathbf{\hat{y}}$, modes $\beta_y$ and $\psi$ would have been coupled. From numerical solutions to Eqs.~(\ref{eq.STT-couples_psi_beta})~and (\ref{eq.STT-couples_psi_beta2}) we can see that if we drive using an STT at the natural frequency of the $\beta_x$-mode we set the coupled $\psi$-mode into exponential growth above a threshold strength of $\eta$, see Fig.~\ref{fig.switch-STT}(b). 

\color{black}
\textbf{Nonadiabatic / field like STT:}
In most situations an adiabatic spin-transfer torque is accompanied by a sizeable nonadiabatic component. To investigate the effects of that we use Rayleigh dissipation function of the form $\mathcal{R'} = \dot{\mathbf{m}}\cdot\mathbf{m}_0$. Considering $\mathbf{m}_0 = (\zeta,\zeta',0)$, the Rayleigh function takes the form:
\begin{equation}
    \mathcal{R'} = \frac{\zeta}{2}(\dot{\beta}_y\beta_x + \dot{\beta}_x\beta_y) - \frac{\zeta'}{2}(\dot{\beta}_x\beta_x - \dot{\beta}_y\beta_y).
    \label{eq.NSTT}
\end{equation}
For $\mathbf{m}_0$ polarized out of the kagome plane, $\mathbf{m}_0 = 2\sqrt{6}(0,0,\zeta_z)$, we have a correction to third order in the field strength:
\begin{equation}
    \mathcal{R'} = \zeta_z \dot{\beta}_y (\beta_x^2 + \beta_y^2).
\end{equation}
As evident this effect appears at a quadratic order in $\bm{\beta}$ for the in-plane polarization, whereas for an out-of-plane polarization the effect appears at a cubic order in fields. Irrespective of the polarization, the nonadiabatic STT does not act on the azimuthal mode $\psi$ to the linear order and hence does not affect the Hall signal or the magnetization in the kagome plane.

The $\beta$ modes are gapped by the DM interaction, see Eq.~(\ref{eq.full-functional}), and will hence undergo oscillations about zero amplitudes, unless significantly larger strains $\epsilon\simeq D$ are applied. If both adiabatic and nonadiabatic (in-plane polarization) STT are present, all modes are coupled, but there will be no change in the AH response unless the STT drives the $\bm{\beta}$ modes near resonant frequencies. The corrections to the equations of motion from this nonadiabatic STT is of the form:
\begin{eqnarray} 
&&\rho' \Ddot{\psi} = \frac{2\delta\epsilon}{J}\sin[2(\psi + \psi_{\epsilon})] + \eta \beta_x - 3\alpha_D\dot{\psi} + 3\eta'(t), \\ \nonumber
&&\frac{2\rho'}{3}\Ddot{\beta}_x = \left(\frac{\epsilon}{2} - 2\sqrt{3}D\right)\beta_x - \eta \psi - \alpha_D\dot{\beta}_x - \frac{\zeta}{2}\beta_y +  \frac{\zeta'}{2}\beta_x,  \\ \nonumber
&&\frac{2\rho'}{3}\Ddot{\beta}_y = -\left(\frac{\epsilon}{2} + 2\sqrt{3}D\right)\beta_y  - \eta - \alpha_D\dot{\beta}_y  - \frac{\zeta}{2}\beta_x- \frac{\zeta'}{2}\beta_y, 
\end{eqnarray}
Here the adiabatic STTs are represented by the out-of-plane component, $\eta'$, and the in-plane component, $\eta$. The out-of-plane component of the nonadiabatic STT only appears at the cubic order in fields. This term is then present only if both time-reversal and inversion symmetries are broken and is expected to be small.
\color{black}

\textbf{Discussion:}
We have demonstrated that the addition of strain can be effectively employed to modulate the STT response of the chiral antiferromagnets Mn$_3$X. The strain converts the symmetry of the system from $C_6$ to $C_2$ in the kagome plane and allows to switch between the six ground states of the system. Each of these six chiral states has a different orientation for the order parameter $\mathbf{K}$ and hence a different AH conductivity tensor explicitly shown in \cite{Liu:2017}.

Thus a controlled protocol for switching between the ground states provides a deterministic way of manipulating the Hall response. Once we manipulate the energy landscape with strain (Fig.~\ref{fig.geometry-modes}) the switching can be affected by oscillating THz magnetic fields or a spin current, using techniques similar to experimentally demonstrated in Ref.~\cite{Disa2020}. We have outlined two switching protocols employing STT: (1) using a pulse of variable width and small amplitude, which switches the Hall angle $\psi$ by $\pm \pi/2$ (Figs.~\ref{fig.osci-mag-switch} and~\ref{fig.osci-strain-STT}), and (2) the one switching $\psi$ by $\pm \pi$, which requires a controlled STT pulse width and amplitude (Fig.~\ref{fig.switch-STT}). 

The theory presented here provides the groundwork for spin-transfer torque based devices in Mn$_3$X. Since the Hall signal is substantial in these materials, a switch in the response should be easily detectable. The additional advantage is that these compounds show the magnetic ordering at room temperatures, and all the way down to cryogenic temperatures for Mn$_3$Ge \cite{mn3ge-hall}.

\begin{acknowledgments}
We are grateful to S.~Duttagupta and S.~Fukami for insightful discussions. S.D. is supported by funding from the Max Planck-UBC-UTokyo Center for Quantum Materials, the Canada First Research Excellence Fund, Quantum Materials and Future Technologies Program, and the Japan Society for the Promotion of Science KAKENHI (Grant No.~JP19H01808). O.A.T. acknowledges the support by the Australian Research Council (Grant No.~DP200101027), the Cooperative Research Project Program at the Research Institute of Electrical Communication, Tohoku University (Japan), and NCMAS grant. 
\end{acknowledgments}

\bibliography{main}

\end{document}